\documentclass[aps,preprint, superscriptaddress,nofootinbib]{revtex4}
\usepackage{amsmath}
\usepackage{graphicx}
\usepackage{subfigure}
\usepackage{makecell}
\usepackage{amssymb}
\usepackage{xcolor}
\usepackage{multirow}
\usepackage{cancel}
\usepackage{color}
\usepackage{epsfig}
\usepackage{ulem}
\usepackage{multirow}
\usepackage{listings}
\usepackage{xcolor}
\usepackage{float}
\usepackage{slashed}

\usepackage{tikz}
\usepackage{pgffor}							

\usepackage[colorlinks,citecolor=blue]{hyperref}
\usepackage{amsmath}
\usepackage{wrapfig}

\newcommand{\be}{\begin{equation}}
\newcommand{\ee}{\end{equation}}
\newcommand{\beq}{\begin{equation}}
\newcommand{\eeq}{\end{equation}}
\newcommand{\bea}{\begin{eqnarray}}
\newcommand{\eea}{\end{eqnarray}}
\newcommand{\besp}{\begin{equation}\begin{split}}
\newcommand{\eesp}{\end{split}\end{equation}}

\newcommand{\nn}{\nonumber}

\newcommand{\Eq}[1]{Eq.~(\ref{#1})}

\newcommand{\Dfbd}{\mathord{\buildrel{\lower3pt\hbox{$\scriptscriptstyle\leftrightarrow$}}\over {D}_{\mu}}}

\def\0{\textbf{0}}
\def\1{\textbf{1}}
\def\2{\textbf{2}}
\def\3{\textbf{3}}
\def\4{\textbf{4}}
\def\5{\textbf{5}}
\def\6{\textbf{6}}
\def\7{\textbf{7}}
\def\8{\textbf{8}}
\def\9{\textbf{9}}

\begin{document}
\title{Cosmological phase transitions,  gravitational waves and self-interacting dark matter in the singlet extension of MSSM}


\author{Wenyu Wang }
\email{wywang@bjut.edu.cn}
\affiliation{Faculty  of  Science, Beijing University of Technology, Beijing, P. R. China}

\author{Ke-Pan Xie}
\email{kepan.xie@unl.edu}
\affiliation{Department of Physics and Astronomy, University of Nebraska, Lincoln, NE 68588, USA}

\author{Wu-Long Xu}
\email{wlxu@emails.bjut.edu.cn}
\affiliation{Faculty  of  Science, Beijing University of Technology, Beijing, P. R. China}

\author{Jin Min Yang}
\email{jmyang@itp.ac.cn}
\affiliation{CAS Key Laboratory of Theoretical Physics, 
Institute of Theoretical Physics, Chinese Academy of Sciences, Beijing 100190, P. R. China}
\affiliation{ School of Physical Sciences, University of Chinese Academy of Sciences, Beijing 100049, P. R. China}

\begin{abstract}
In the minimal supersymmetric standard model (MSSM) extended by a singlet superfield, when the coupling between the singlet sector and the MSSM sector is tiny, the singlet sector can be a quasi dark sector with supersymmetry (SUSY). We investigate the cosmological phenomena in this scenario and obtain the following observations:
(i) In the parameter space solving the small cosmological scale anomalies via self-interacting singlino dark matter (SIDM),  a first-order phase transition (FOPT) can readily happen but requires rather light dark matter below MeV; 
(ii) The corresponding parameter space indicated by FOPT and SIDM can be partially covered by detecting the phase-transition gravitational waves (GWs) 
at the near-future projects, such as LISA, TianQin and Taiji. Therefore, the recently developed GW astronomy could be a novel probe to such a SUSY scenario.
\end{abstract}

\maketitle

\tableofcontents

\section{Introduction}

Dark matter (DM) and gravitational wave (GW) are hot topics in the frontier of both theoretical and experimental physics studies. The DM is indicated from lots of astrophysical observations, which is non-luminous and distributes all over the Universe with a density approximately five times of the ordinary matter~\cite{Young:2016ala}. Usually the DM is considered as a fundamental particle which is weakly interacting with the Standard Model (SM) particles to give the correct relic density, i.e., the so-called  WIMP (Weakly Interacting Massive Particle) miracle~\cite{Dodelson:2003ft}. 
As a mystery of the fundamental structure of matters, the DM issue is on the list of the physics beyond the SM and needs to be deeply studied. On the other hand, the GW, which was a simple prediction of classical general relativity long time ago, has been measured very recently due to the great improvements in the measurements of space and time. Superficially these two topics are unrelated, which, however, become correlated in case of a first-order cosmological phase transition (FOPT)~\cite{Linde:1981zj}.   

In the particle physics side, the spontaneous symmetry breaking of the Higgs field, which is a fundamental field of phase transition, gives the masses of particles. However, there is no DM candidate in the SM. And the SM suffers from both aesthetical judgements and experimental observations. The hierarchy problem, the strong CP problem and the Grand Unification Theory (GUT) require new content of particles or new symmetries; in addition to DM, the neutrino oscillation implies finite mass of neutrinos. Although finite neutrino mass terms can be introduced conventionally, in the form of Higgs terms without a significant modification of the theory, this makes the SM to appear rather contrived. Thus the physics beyond the SM with an upgraded phase transition usually has to solve the hierarchy problem, realize GUT, and give candidates of the DM. In the GW side, if the potential of a scalar field has two different energy minima, the FOPT can happen via quantum tunneling between the two vacua separated by a barrier~\cite{Coleman:1977py,Callan:1977pt}. In the early Universe, as temperature was dropping, this quantum tunneling may also occur in the Higgs field. During the FOPT, the collision of vacuum bubbles, sound waves in the plasma and the magneto-hydrodynamics turbulence could radiate GWs that may be detected by the future space-based LISA~\cite{LISA:2017pwj}, TianQin~\cite{TianQin:2015yph} or Taiji~\cite{Ruan:2018tsw} experiments. Thus, for an unified description of the underlying fundamental theory, the results of phase transition should be checked from both sides since phase transition in fact gives a platform for both fundamental particles and gravitational phenomena, or the detection of phase transition should give constraints on fundamental particle models.

The Monte Carlo simulations demonstrate that the electroweak phase transition in the SM is a crossover~\cite{Kajantie:1996mn,Fodor:1994sj,Kajantie:1993ag}. It shows that the extension of the SM is needed for a FOPT~\cite{Espinosa:2011ax,Athron:2019teq,Apreda:2001us,Hasegawa:2019amx,Bian:2019kmg,Xie:2020bkl,Su:2020pjw,Ghosh:2020ipy,Guo:2021qcq,Goncalves:2021egx}. Among various new physics models, low energy supersymmetry (SUSY) is most widely studied since it can elegantly solve both aesthetical and experimental problems of the SM~\cite{Wess:1974tw, Sohnius:1985qm, Sakai:1981gr,Dimopoulos:1981zb} (for review reviews, see, e.g., Refs.~\cite{Baer:2020kwz,Wang:2022rfd}).
However, the FOPT can be hardly realized in the minimal supersymmetric standard model (MSSM) because the needed large barrier of the SM Higgs potential requires light stops (superpartners of top quark) not allowed by current search results at colliders. This problem can be easily solved by adding a singlet superfield which can have 
tree-level barrier freely. The FOPT of such singlet extension of MSSM, which is called NMSSM (GNMSSM) with (without) $Z_3$ symmetry \cite{Ellwanger:2009dp,Cao:2012fz}, has been studied intensively in the literature~\cite{Huang:2014ifa,Bi:2015qva,Bian:2017wfv,Athron:2019teq,Chatterjee:2022pxf}. In these studies, the phase transition is always considered at the electroweak scale and the singlet couples with Higgs doublets to play the role of spontaneous symmetry breaking.  Generally speaking, the coupling between the singlet and Higgs doublets, which is the source of electroweak phase transition, is not the essential part in theoretical models. The phase transition of singlet itself can have more advantages compared with the mixing with the doublets. The most promising point is that the singlet superfield can play a very subtle role in the scenario of light DM and self-interacting DM (SIDM) favored by the observation of small structures of the Universe. The singlino can be a light DM candidate, annihilating to the singlet scalars to give correct relic abundance,  interchanging with singlet to give appropriate self-interaction of singlino DM and solve the small structure problem of the WIMP. The most important thing is that the singlino DM with tiny coupling to the doublets can easily escape the current stringent direct detection limits of the DM. Thus, this scenario has charming features in DM physics, which can realize SIDM scenario and solve the small cosmological scale anomalies~\cite{Wang:2014kja,Zhu:2021pad}. 

With these advantages in fundamental particle physics, the scenario of a singlet superfield with tiny couplings to the MSSM sector should be checked from the cosmological phase transition and the detection of GWs. In details, we should check: whether the FOPT can be maintained, whether the surviving parameter space can be probed by the detection of the GWs, or whether the GW detection
can cover the parameter space of SIDM. In this work we examine these issues. This work is organized as follows. In Section~\ref{sec2} we show the content of the SUSY singlet model and show how the small structure problem is solved. In Section~\ref{sec3}, the FOPT and GWs are discussed. In Section~\ref{sec4} the numerical results are shown. The conclusion is given in Section~\ref{sec5}.

\section{Self-interacting singlino dark matter}\label{sec2}

As addressed in the preceding section, the non-luminous DM is a necessary component in the Universe. The standard cosmological model $\Lambda$CDM, which assumes cold and collisionless DM, has achieved remarkable success in describing the Universe at large scales greater than $\cal O$ (Mpc) today. However, this model encounters lots of crises on the small-scale Universe at which the structure formation becomes strongly non-linear and the $N$-body simulation is the standard tool in this regime. The numerical results showed that the weakly interacting DM cannot form the rich structures observed in the Universe. Such discrepancies are called the small structure problems, including the core-cusp problem~\cite{Navarro:1996gj,Moore:1999gc}, the diversity problem~\cite{KuziodeNaray:2009oon,Bullock:1999he,Oman:2015xda}, the missing satellites problem~\cite{Klypin:1999uc,Kauffmann:1993gv,Zavala:2009ms} and the too-big-to-fail problem~\cite{Boylan-Kolchin:2011lmk,Tollerud:2014zha,Garrison-Kimmel:2014vqa}.

Though in the literature there are some debates about the robustness of these problems, they may indeed indicate some shortcomings of CDM in describing the small structures of the Universe. One way to solve these problems is to propose warm DM, which,   however, is not consistent with some other observations~\cite{Irsic:2017ixq,Viel:2013fqw,Menci:2016eui} and gives too small core sizes to solve the core-cusp problem~\cite{Maccio:2012qf}.

Another promising alternative approach to solve the small structure problems is to introduce self-interacting dark matter (SIDM)~\cite{Spergel:1999mh}. For SIDM the elastic cross section is velocity-dependent and thus the light mediator of SIDM can give sufficiently large cross sections to satisfy the small structure requirements, while the annihilation can still maintain weak to give a correct freezing-out relic density in the early Universe.

A detailed study (shown in the proceeding section) of SIDM shows that the mediators or the force carriers prefer at about several MeV or even lighter. Such a hierarchy between the mediator and electroweak scale makes it very difficult to connect with the electroweak scale physics in the scheme of the  new physics beyond SM. Further more, as shown in Ref.~\cite{Wang:2014kja,Elor:2021swj}, if the light mediator couples to the nucleons, the correlation between the DM annihilation rate and DM-nucleon spin independent (SI) cross section has almost excluded the scenario. Thus from the point view of SIDM, the introduction of a singlet is not necessarily connected to the electroweak sector.

With a tiny connection to electroweak sector, a singlet in the SUSY framework can be a perfect building block for SIDM. The singlino composes the DM, annihilating to the singlet bosons to give the correct relic abundance, and interacting by exchanging singlet bosons to give large scattering elastic cross sections in case of small velocities for the small structure cosmological scales. In the following, we will show the details of such a singlet SUSY model.

\begin{figure}[H]
	\centering
	\includegraphics[width=12cm]{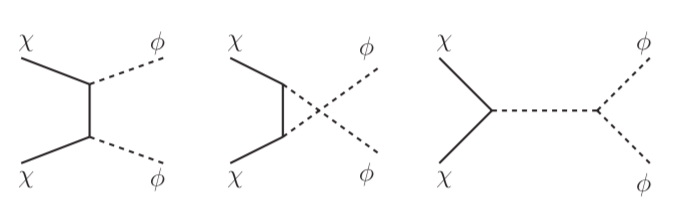}
	\caption{ Feynman diagrams of singlino annihilation  to the salars, with $\phi=H$, $A$.}
	\label{fig1}
\end{figure}

The superpotential of a singlet in the most general form is
\begin{equation}\label{GMSSM superpotential}
W=\eta\hat{S}+\frac{1}{2}\mu_s\hat{S}^2+\frac{1}{3}\kappa\hat{S}^3,
\end{equation}
where $\hat S$ is the singlet superfield, $\eta$ and $\mu_s$ are parameters with mass dimension $m^2$ and $m$ respectively, and $\kappa$ is a dimensionless coupling strength of the field. The fermion part of $\hat S$ gives a Majorana fermion singlino, while the scalar part $S$ gives one scalar and one pseudo-scalar. This is actually the modified Wess-Zumino model, with simply degenerate mass spectrum. A physically allowed model needs soft breaking terms
\begin{eqnarray}
  V_{\rm soft}
=m_{s}^{2}|S|^{2}+(C_{\eta}\eta S+\frac{1}{2}B_{s}\mu_{s}S^{2}+\frac{1}{3}\kappa A_{k}S^{3}+{\rm h.c.}),
\end{eqnarray}
and then the potential becomes 
\begin{equation}
V=V_F+V_{\rm soft}
=|\kappa S^{2}|^{2}+m_{s}^{2}|S|^{2}+\left(C_{\eta}\eta S
+\frac{1}{2}B_{s}\mu_{s}S^{2}+\frac{1}{3}\kappa A_{k}S^{3}+\kappa\mu_{s}S^{2}S^{*}
+{\rm h.c.}\right).\label{singlet_potential}
\end{equation}
Note that some of the parameters are absorbed by the redefinition of the soft parameters, giving a more economic potential. We can see that this potential can easily spontaneously break to get the vacuum expectation value (VEV) $v_s$. Thus we denote
\begin{eqnarray}
  S =v_s+ \frac{1}{\sqrt{2}}(H+iA).
\end{eqnarray}
The tadpole vanishes in the condition of 
\begin{eqnarray}\label{eqms}
 m_{s}^{2}=-2 v_s^{2}\kappa^{2}-v_s\kappa A_{\kappa}
-\frac{\eta C_{\eta}}{v_s}-(3v_s\kappa+B_{s})\mu_{s}.
\end{eqnarray}
After a short calculation, we can get the masses of the singlino $\chi$, CP-even scalar $H$ and CP-odd pseudo-scalar $A$:
\be\label{particle_mass}\begin{split}
m_{\chi}&=~\mu_{s}+2\kappa v_s,\\
m_{H}^{2}&=~6v_s^{2}\kappa^{2}+2v_s\kappa A_{\kappa}+m_{s}^{2}+6v_s\kappa\mu_{s}+B_{s}\mu_{s},\\
m_{A}^{2}&=~2v_s^{2}\kappa^{2}-2v_s\kappa A_{\kappa}+m_{s}^{2}+2v_s\kappa\mu_{s}-B_{s}\mu_{s}.
\end{split}\ee
The corresponding Feynman rules are shown in the appendix. Here we just mention that singlino $\chi$ can have sizable self-interaction by exchanging a light scalar, and the annihilation to the scalars can give a proper relic abundance of singlino DM. The corresponding Feynman diagrams of the annihilation are shown in Fig.~\ref{fig1}.

Note that, in this singlet SUSY model, the DM annihilation has more channels than in the simple SIDM model with only one DM and one mediator, in which the DM can only annihilate to the scalar via $t$- and $u$-channels (the first two diagrams in Fig. \ref{fig1}). In our model, the scalars have self-couplings, thus the annihilation can have $s$-channel (the third diagram in Fig. \ref{fig1}). This will make the parameter space different from the simple SIDM model. Nevertheless, the self-interactions of scalars are in fact derived from the scalar potential \Eq{singlet_potential}, which  is the starting point of the FOPT for the detection of the GWs, and thus detailed studies are needed to check the connection between annihilation, self-interaction and GW detection.

After the derivation of the basic model, an important issue is the parameter definition. Though we can use the soft parameters such as $\mu$ and $C_\eta$ to study the model, the physical implications of these parameters are obscure. Thus we chose $m_{\chi}$, $m_H$, $m_A$, $\alpha_{\chi}$, $A_{\kappa}$, $v_s$ as the parameters of the model, which are masses of the particles,  the coupling strength, the trilinear term and the scalar VEV, respectively. Note that we chose $\alpha_{\chi}=4\kappa^2/\pi$ for a more convenient definition of the coupling strength. Then, the scalar potential \Eq{singlet_potential} can be obtained from \Eq{eqms} and we have the following relations
\be\label{reparameyer}\begin{split}
\kappa=&~\frac{\sqrt{\pi\alpha_{\chi}}}{2},~~~~
\mu_{s}=m_{\chi}-2\kappa v_s,\\
\eta C_{\eta}=&~\frac{1}{2}v_s(-m_{H}^{2}+(A_{\kappa}+3m_{\chi})\sqrt{\pi\alpha_{\chi}}v_s-\pi\alpha_{\chi}
v_s^{2}),\\
B_{s}=&~\frac{m_{H}^{2}-m_{A}^{2}+v_s(-2m_{\chi}+\sqrt{\pi\alpha_{\chi}}(-4A_{\kappa}-3m_{\chi}+2v_s)+\pi\alpha_{\chi}v_s)}{4(m_{\chi}-\sqrt{\pi\alpha_{\chi}}v_s)}.
\end{split}\ee
These parameters are basic inputs for the scalar potential.  Before going to the detailed study of FOPT, we should note that our scalar potential is not a simple construction from the scalar field with specific gauge symmetry, instead it is just a general construction of a singlet superfield, or namely, the soft breaking Wess-Zumino SUSY model. Thus the model has particular meaning on the study of SUSY, a mystery symmetry of the space and time of our Universe.

\section{Cosmic FOPT and gravitational waves}\label{sec3}

Cosmological FOPT is the decay of the Universe between the false and true vacua, which are separated by a potential barrier~\cite{Linde:1981zj}. In our case, the FOPT scalar field is the CP-even background scalar field $h$, whose potential receives the quantum and thermal corrections from the CP-even scalar itself, the CP-odd scalar and the singlino loops. A detailed study of FOPT and detection of the GWs are discussed in the following.

\subsection{First-order phase transition}

We work under the CP conservation assumption such that only the real part of the complex field $S=(H+iA)/\sqrt{2}$ gains a VEV, and hence we only consider the CP-even background field $h$. At zero temperature, the effective potential at the tree level is
\begin{eqnarray}\label{zerotemperaturepotential}
V_{0}(h)=\kappa^{2}h^{4}+m_{s}^{2}h^{2}+2\left(C_{\eta}\eta h+\frac{1}{2}B_{s}\mu_{s}h^{2}+\frac{1}{3}\kappa A_{k}h^{3}+\kappa\mu_{s}h^{3}\right).
\end{eqnarray}
At one-loop level, the potential receives the Coleman-Weinberg corrections
\begin{eqnarray}\label{oneloopcorrectionzero}
V_{1}^{T=0}(h)
&=&\frac{1}{64\pi^{2}}m_a(h)^{4}\left[\log\frac{m_{a}^{2}(h)}{\mu^{2}}-\frac{3}{2}\right]+\frac{1}{64\pi^{2}}m_h^{4}(h)\left[\log\frac{m_{h}^{2}(h)}{\mu^{2}}-\frac{3}{2}\right],\nonumber\\
&&\,-\frac{1}{32\pi^{2}}m_{\chi}^{4}(h)\left[\log\frac{m_{\chi}^{2}(h)}{\mu^{2}}-\frac{3}{2}\right],
\end{eqnarray}
where we use the $\overline{\rm MS}$ subtraction scheme and take the renormalization scale as $\mu=246~{\rm GeV}$. Note that there is a SUSY limit in the potential. $V_{1}^{T=0}(h)$ will approach zero when the masses of $\chi$, $H$, $A$ approach to degenerate, which is an interesting point in SUSY theory. The field-dependent masses are obtained by replacing $v_s$ with $h$ in \Eq{particle_mass}. To remain the tree level VEV and mass relations, we also add the following counter terms
\begin{eqnarray}\label{zerotemperaturepotential1}
\delta V_1^{T=0}(h)=\delta m_{s}^{2}h^{2}+2\left(\delta C_{\eta}\eta h+\frac{1}{2}B_{s}\delta\mu_{s}h^{2}+\frac{1}{3}\delta\kappa A_{k}h^{3}\right),
\end{eqnarray}
and choose the renormalization condition as
\begin{eqnarray}
\frac{d}{dh}\left(V_{1}^{T=0}+\delta V_1^{T=0}\right)\Big|_{h=v_s}=0,\quad \frac{d^2}{dh^2}\left(V_{1}^{T=0}+\delta V_1^{T=0}\right)\Big|_{h=v_s}=0.
\end{eqnarray}

At finite temperature, the potential receives one-loop thermal integral corrections 
\begin{eqnarray}\label{oneloopcorrectiontemperaturecorrection}
&&V_1^{T\neq0}(h,T)=\frac{T^{4}}{2\pi^{2}}J_{B}\left(\frac{m_{h}^{2}(h)}{T^{2}}\right)+\frac{T^{4}}{2\pi^{2}}J_{B}\left(\frac{m_{a}^{2}(h)}{T^{2}}\right)-4\frac{T^{4}}{2\pi^{2}}J_{F}\left(\frac{m_\chi^{2}(h)}{T^{2}}\right),
\end{eqnarray}	
where
\begin{equation}
J_{B/F}(y)=\int_{0}^{\infty}dxx^{2}\log\left(1\mp e^{-\sqrt{x^{2}+y}}\right).
\end{equation}
We also include the daisy resummation contribution
\be
V_{\rm daisy}(h)=-\frac{T}{12\pi}\left[\left(m_h^2(h)+\Pi_hT^2\right)^{3/2}+\left(m_a^2(h)+\Pi_aT^2\right)^{3/2}-m_h^3(h)-m_a^3(h)\right],
\ee
where $\Pi_h=\kappa^{2}T^{2}$, $\Pi_a=\kappa^{2}T^{2}$.

The total effective potential up to one-loop can be written as
\begin{equation}\label{effective potential}
V_{\rm eff}(h,T)=V_0(h)+V_1^{T=0}(h)+\delta V_1^{T=0}(h)+V_1^{T\neq 0}(h,T)+V_{\rm daisy}(h,T).
\end{equation}
The temperature that $V_{\rm eff}(h,T)$ has two degenerate vacua is defined as the critical temperature $T_c$. Below $T_c$, the true vacuum is energetically preferred and the Universe will decay from the false vacuum to the true vacuum. The decay rate per unit volume is given by~\cite{Linde:1981zj}
\be
\Gamma\sim T^4\left(\frac{S_3(T)}{2\pi T}\right)^{3/2}e^{-S_3(T)/T},
\ee
where the Euclid action $S_3$ is
\be\label{S_3}
S_3=\int_0^\infty 4\pi r^2dr\left[\frac12\left(\frac{d\hat h}{dr}\right)^2+V_{\rm eff}(\hat h,T)\right],
\ee
with $\hat h(r)$ being the $O(3)$-symmetric bounce solution determined by
\be
\frac{d^2\hat h}{dr^2}+\frac{2}{r}\frac{d\hat h}{dr}=\frac{\partial }{\partial\hat h}V_{\rm eff}(\hat h,T),\quad
\lim_{r\to\infty}\hat h=0,\quad\frac{d\hat h}{dr}\Big|_{r=0}=0.
\ee
The probability that a bubble nucleates inside a Hubble volume is
\be
N(T)=\int_T^{T_c}\frac{dT'}{T'}\frac{\Gamma(T')}{H^4(T')},
\ee
where $H(T)$ is the Hubble constant. The temperature satisfying $N(T_n)\sim1$ is called the nucleation temperature $T_n$. This is resolved by 
\be\label{OrigS3T}
\frac{S_3}{T_n}\simeq \ln\left[\frac{1}{4}\left(\frac{90}{8\pi^3 g_*}\right)^2\right]+4 \ln \left[\frac{M_{\rm Pl}}{T_n}\right],
\ee
where $M_{\rm Pl}=1.22\times10^{19}$ GeV is the Planck scale, and $g_*$ is the number of relativistic degrees of freedom. Here, we can take $T_n$ of right hand side of \Eq{OrigS3T} as  $T_c$ approximately because of the logarithmic function. In this article,  we use the public package {\tt cosmoTransition}~\cite{Wainwright:2011kj}\footnote{See Refs.~\cite{Athron:2020sbe,Athron:2019nbd} for other packages for calculating the FOPT.} to calculate the bounce solution and determine $T_n$.

\subsection{Gravitational waves}

Next, we show the details of the radiation of the GWs. The volume fraction of the false vacuum of the Universe is given by~\cite{Guth:1981uk}
\be
p(T)=e^{-I(T)},\quad I(T)=\frac{4\pi}{3}\int_{T}^{T_c}dT'\frac{\Gamma(T')}{T'^4H(T')}\left[\int_{T}^{T'}d\tilde T\frac{v_b}{H(\tilde T)}\right]^3,
\ee
which decreases to zero as the FOPT completes. Here $v_b$ is the bubble expansion velocity relative to the plasma at finite distance. The temperature that the true vacuum bubbles form an infinite connected cluster is called the percolation temperature $T_p$. Numerical simulation shows that $p(T_p)=0.71$~\cite{rintoul1997precise}. $T_p$ is the characteristic temperature 
for calculating the FOPT GWs~\cite{Megevand:2016lpr,Kobakhidze:2017mru,Ellis:2018mja,Ellis:2020awk,Wang:2020jrd}.

A FOPT generates stochastic GWs via bubble collision, sound waves and turbulence of the magneto-hydrodynamics (MHD) in the plasma. Because of the friction in the plasma-wall system, the expansion of bubble wall will accelerate for only a short time and then reach the terminal velocity $v_b$. Therefore, only a small fraction of FOPT energy is stored in the wall and the bubble collision contribution is negligible. Instead, most energy is in the fluid shells surrounding the wall, making the sound waves the dominant contribution~\cite{Ellis:2018mja}. The MHD turbulence contributes a sub-leading source to the GWs. Defining the GW spectrum today as
\be\label{GW}
\Omega_{\rm GW}(f)=\frac{1}{\rho_c}\frac{\rho_{\rm GW}}{d\ln f},
\ee
where $f$ is the frequency, $\rho_{\rm GW}$ is the GW energy density and $\rho_c$ is the critical energy density of the present Universe, we have\footnote{Here $h$ is $H_0/(100~{\rm km/s/Mpc})$, not to be confused with the background field $h$.}
\be\label{GWsignal}
\Omega_{\rm GW}h^{2}\simeq\Omega_{\rm sw}h^{2}+\Omega_{\rm turb}h^{2},
\ee
where the sound wave contribution $\Omega_{\rm sw}$ and turbulence contribution $\Omega_{\rm turb}$ can be expressed as the numerical functions of two parameters from the FOPT profile~\cite{Grojean:2006bp,Caprini:2015zlo,Caprini:2019egz}: i) $\alpha$, defined as the ratio of the latent heat in the FOPT to the radiation energy density of the Universe; and ii) $\beta/H_p$, the inverse ratio of duration of FOPT and the Hubble time. Quantitatively, their definitions are
\be
\alpha=\frac{1}{\rho_R(T_p)}\left(T\frac{\partial\Delta V_{\rm eff}}{\partial T}-\Delta V_{\rm eff}\right)\Big|_{T_p};\quad \frac{\beta}{H_p}=T_p\frac{d(S_{3}/T)}{dT}\Big|_{T_p},
\ee
where $\Delta V_{\rm eff}$ is the free energy difference between true and false vacua and $\rho_R(T)=g_*\pi^2T^4/30$ is the radiation energy density (with $g_*$ being the number of relativistic degrees of freedom). We adopt $T_p\approx T_n$ for the GW calculation, as the supercooling effect is not strong in our scenario, and the characteristic temperatures are very close. Because the coupling between dark sector and SM is extremely weak, after reheating stage, those two parts do not have any chance exchanging energy and injecting entropy until the dark sector phase transition. Thus, the total number of relativistic degrees of freedom is $g_*=g^{\rm SM}_*+g^d_*(\frac{T}{T_{\rm SM}})$ when the dark sector temperature is $T$ and the SM temperature is $T_{\rm SM}$. Here we take $g^d_*=3.75$ counting the degrees of freedom from $\chi$, $H$ and $A$. This latent heat changes the dark sector temperature form $T_p$ to $T_f$. The result is written as 
\be \label{Tf}
\frac{\pi^{2}g_{*}^{d}}{30}T_{f}^{4}=\frac{\pi^{2}g_{*}^{d}}{30}T_{p}^{4}+\rho_{\rm dvac}.
\ee 
The parameter $\alpha$ and $\frac{\beta}{H_p}$ can be used to characterize the strength of the FOPT.

Numerically, the sound wave contribution reads
\begin{eqnarray}
&&h^{2}\Omega_{\rm sw}(f)=2.65\times10^{-6}(H_p\tau_{\rm sw})\left(\frac{H_{p}}{\beta}\right)\left(\frac{\kappa_{v}\alpha}{1+\alpha}\right)^{2}\left(\frac{100}{g_{*}}\right)^{\frac{1}{3}}v_bS_{\rm sw}(f),\\
&&S_{\rm sw}(f)=\left(\frac{f}{f_{\rm sw}}\right)^{3}\left(\frac{7}{4+3(\frac{f}{f_{\rm sw}})^{2}}\right)^{\frac{7}{2}},\\
&&f_{\rm sw}=1.9\times10^{-2}~{\rm mHz}\frac{1}{v_b}\left(\frac{\beta}{H_{p}}\right)\left(\frac{T_{p}}{100~{\rm GeV}}\right)\left(\frac{g_{*}}{100}\right)^{\frac{1}{6}},
\end{eqnarray}
where $\tau_{\rm sw}$ is the duration of the sound wave source~\cite{Ellis:2018mja,Guo:2020grp}, and $\kappa_v$ is the fraction of vacuum energy distributed in the bubble kinetic energy and can be found in Ref.~\cite{Espinosa:2010hh}.
The turbulence contribution is 
\begin{eqnarray}\label{turbulence}
&&h^{2}\Omega_{{\rm turb}}(f)=3.35\times10^{-4}\left(\frac{H_{p}}{\beta}\right)\left(\frac{\kappa_{\rm turb}\alpha}{1+\alpha}\right)^{2}\left(\frac{100}{g_{*}}\right)^{\frac{1}{3}}v_bS_{\rm sturb}(f),\\
&&S_{\rm turb}(f)=\frac{(\frac{f}{f_{\rm turb}})^{3}}{[1+(\frac{f}{f_{\rm turb}})]^{\frac{11}{3}}\left(1+\frac{8\pi f}{h_{p}}\right)},\\
&&f_{\rm sw}=2.7\times10^{-2}~{\rm mHz}\frac{1}{v_b}\left(\frac{\beta}{H_{p}}\right)\left(\frac{T_{p}}{100~{\rm GeV}}\right)\left(\frac{g_{*}}{100}\right)^{\frac{1}{6}},\\
&&\kappa_{\rm turb}=\epsilon\kappa_{v},
\end{eqnarray}
where $\kappa_{\rm turb}$ denotes the fraction of fraction of vacuum energy distributed in the turbulence and $H_p$ is the Hubble parameter at $T_p$.

\section{Numerical results}\label{sec4}

Before performing specific numerical calculation, we briefly review the calculation of DM relic density and SIDM cross section. The Feynman diagrams of the annihilation are shown in Fig.~\ref{fig1} and the  annihilation cross section is~\cite{Drees:1992am, Jungman:1995df}
\beq
\sigma v =\frac{1}{4}\frac{\bar\beta_f}{8\pi sS}\left[
|A(^1S_0)|^2+\frac{1}{3}\left(|A(^3P_0)|^2+|A(^3P_1)|^2\right)+|A(^3P_2)|^2\right],
\eeq
where $S$ is the symmetry factor, amplitudes $A(^1S_0),~A(^3P_0),~A(^3P_1),~A(^3P_2)$
are the contributions from different spin states of DM which are shown in the appendix, and $\bar\beta_f$ is given by
\bea
\bar\beta_f=\sqrt{1-2(m_X^2+m_Y^2)/s + (m_X^2+m_Y^2)^2/s^2}~,
\eea
with $X$, $Y$ being the final states.  

For the SIDM, we use the cross sections defined in Ref. ~\cite{PhysRevA.60.2118},\footnote{We assume a instant freeze-out such that the annihilation cross section can be adopted as the WIMP one~\cite{Vanderheyden:2021gpj}.} the transfer cross section $\sigma_T$ and the viscosity (or conductivity) cross section $\sigma_V$:~\cite{Tulin:2012wi,Tulin:2013teo,Ko:2014nha}
\beq
\sigma_T =  \int d\Omega \, (1-\cos\theta) \,\frac{d\sigma}{d\Omega}  \, , \qquad  \label{sigmaT}
\sigma_V =  \int d \Omega \, \sin^2 \theta \, \frac{d\sigma}{d\Omega} \, .
\eeq
$\sigma_T$ is for the estimation of the Dirac DM, while $\sigma_T$ is for the Majorana DM. Since the singlino $\chi$ is a Majorana fermion, 
the viscosity cross section is defined with two variables:
\bea
\frac{d\sigma_{VS}}{d\Omega} &=& \left|f(\theta)+f(\pi-\theta)\right|^2 = \frac{1}{k^2}\left|
\sum_{\ell (\mbox{\tiny\rm EVEN\  number})}^{\infty} (2\ell + 1) (\exp(2i\delta_l)-1)P_\ell(\cos\theta)\right|^2,
\label{sigmaVSsum}\\
\frac{d\sigma_{VA}}{d\Omega} &=& \left|f(\theta)-f(\pi-\theta)\right|^2 = \frac{1}{k^2}\left|
\sum_{\ell (\mbox{\tiny\rm ODD\ number})}^{\infty} (2\ell + 1) (\exp(2i\delta_l)-1)P_\ell(\cos\theta)\right|^2.
\label{sigmaVAsum}
\eea
Using the orthogonality relation for the Legendre polynomials, we can get
\bea
\frac{\sigma_{VS}k^2}{4\pi} &=& \sum_{\ell(\mbox{\tiny EVEN\ number})}^{\infty}4\sin^{2}(\delta_{\ell+2}
-\delta_{\ell})(\ell+1)(\ell+2)/(2\ell+3) ,\label{sigmaVS}\\
\frac{\sigma_{VA}k^2}{4\pi} &=& \sum_{\ell(\mbox{\tiny ODD\ number})}^{\infty}4\sin^{2}(\delta_{\ell+2}
-\delta_{\ell})(\ell+1)(\ell+2)/(2\ell+3) . \label{sigmaVA1}
\eea
The phase shift $\delta_{\ell}$ must be computed by solving the Schr\"{o}dinger equation
\be
\frac{1}{r^2} \frac{d}{dr} \Big( r^2 \frac{d R_{\ell}}{dr} \Big)
+ \Big( k^2 - \frac{\ell (\ell + 1)}{r^2} - 2m_r V(r) \Big) R_\ell = 0,
\ee
directly with partial wave expansion method. The total wave function of the  spin-1/2 fermionic DM must be antisymmetric with respect to the exchange of two identical particles. Then the spatial wave function should be symmetric when the total spin is 0 (singlet)
while the spatial wave function should be antisymmetric when the total spin is 1 (triplet). Then we consider that the DM scatters with random orientations, thus the triplet is three times as likely as the singlet and the average cross section is
\bea
\sigma_{V} =\frac{1}{4}\sigma_{VS}+\frac{3}{4} \sigma_{VA}\; .\label{sigmaVa;;}
\eea
We need to do a specific numerical calculation to check the existence of parameter space which can satisfy the constraint of relic density, realize the SIDM scenario and give detectable GWs. 

\begin{figure}[h]
	\centering
\includegraphics[width=10cm]{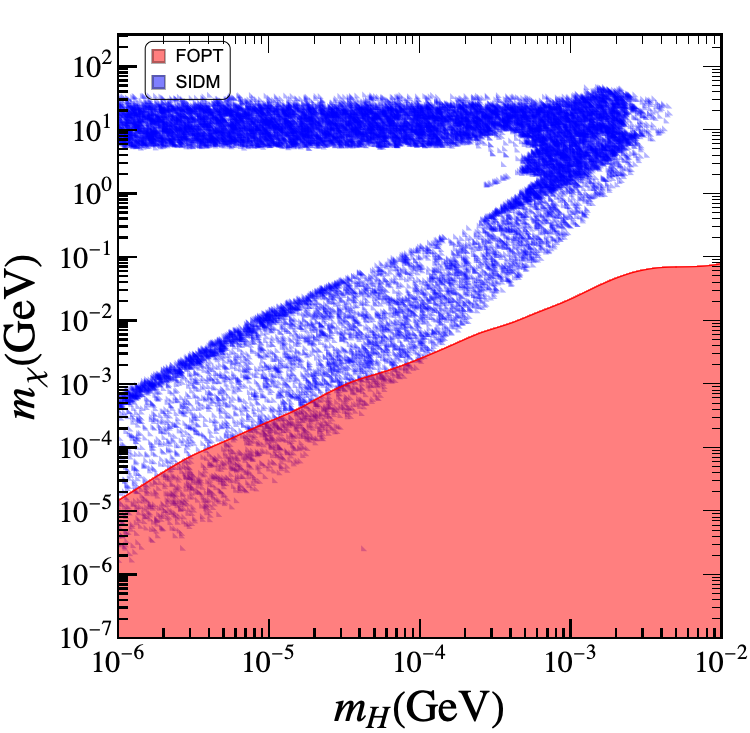}
	\caption{The blue points are 
	allowed by DM relic density and the small scale structure via SIDM; 
	the red region can give a FOPT.
	}
	\label{fig2}
\end{figure}

After all preparations are completed, the parameters $m_{\chi}$, $m_H$, $m_A$, $\kappa$, $A_{\kappa}$ are chosen in ranges of \begin{equation}\label{parameterspace}
\begin{split}
&10^{-6}~{\rm GeV}<m_{\chi}<10^{4}~{\rm GeV},\quad  10^{-6}~{\rm GeV}<m_H<10^{4}~{\rm GeV}, \\
&10^{-6}~{\rm GeV}<m_A<10^{4}~{\rm GeV},\quad
10^{-6}<\kappa<1,\quad -10^{5}~{\rm GeV}<A_{\kappa}<0.
\end{split}
\end{equation} 
According to the experimental constraints, the favourable relic density is $(0.107,0.131)$. For the SIDM cross section $\sigma/m_{\chi}$, Ref.~\cite{Tulin:2017ara} 
shows the different value in the different small scale structure problems. From a pure theoretical point of view, it is sufficient to 
choose $\sigma/m_{\chi}$ in the range of $(0.1, 10~\rm{cm^2/g})$ and the characteristic velocity is $200 ~\rm{km/s}$. 
The parameter space allowed by DM relic density and the small scale structure is shown in  Fig.~\ref{fig2} as the blue points. We can see that the allowed mass of SIDM for the small scale structure can be from keV to tens of GeV, while the mass of dark scalar is below a few MeV. 

For the calculation of FOPT, there is one more free parameter, namely the VEV $v_s$. By varying $v_s\in[10^{-6}~{\rm GeV},10^{4}~{\rm GeV}]$, we obtain the parameter space giving a FOPT, shown as the red region in Fig.~\ref{fig2}. The intersection region between blue points and the red region in Fig.~\ref{fig2} is the  surviving parameter space which can give a correct relic density and realize FOPT and SIDM. It should be noted that in recent studies the filtered bubble can reflect the DM particles because of the energy defect between the value of false 
vacuum and real vacuum of the scalar in which Fermi-ball or Q-ball can be formed~\cite{Krylov:2013qe,Hong:2020est,Kawana:2021tde}. But in this SUSY model, the DM mass gap between the two vacua is small. Thus, the DM particle can pass the bubble wall safely. Then this FOPT can not affect the DM generation and the relic density of DM can be obtained via the standard freeze-out mechanism.
From Fig.~\ref{fig2}, we can see that in the survived region the SIDM is lighter than 1 MeV and the dark scalar is lighter than 100 KeV.

\begin{figure}[h]
	\centering
	\includegraphics[width=5.25cm]{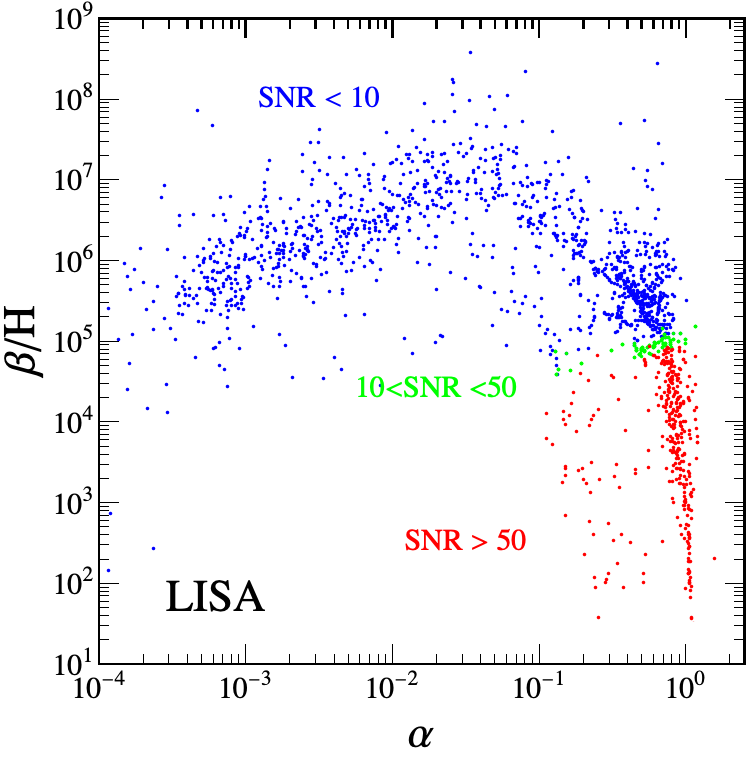}
	\includegraphics[width=5.25cm]{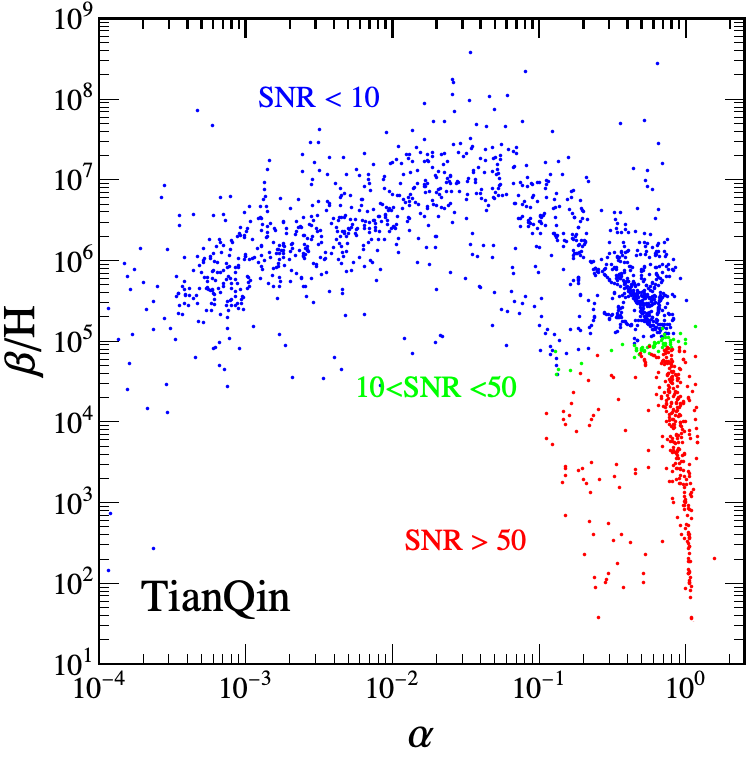}
	\includegraphics[width=5.25cm]{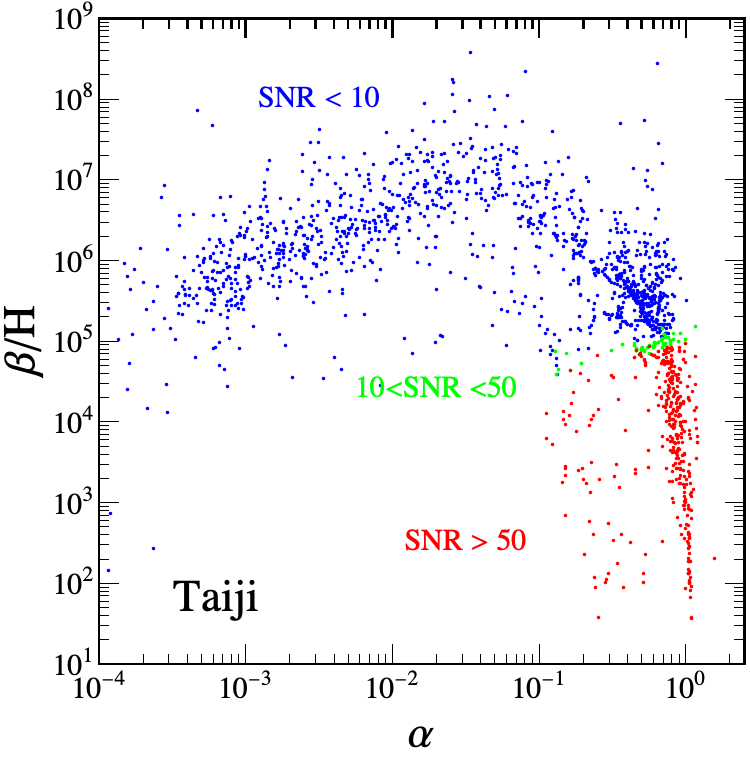}
	\caption{The SNRs at different near-future detectors (left: LISA, middle: TianQin, right: Taiji), with the blue points for SNR $<$ 10,
	the green points for $10<$ SNR $<$ 50 and the red points for SNR $>$ 50.
	The data-taking duration is adopted as four years.}
	\label{fig3}
\end{figure}

According to Refs.~\cite{Marfatia:2021twj,DiBari:2021dri}, the effective number of extra neutrino species contributed by the SUSY dark sector $\Delta N_{\rm eff}\propto \frac{T_f}{T_{\rm SM}}$. Because the phase transition belongs to low temperature FOPT, the constraints from the epochs of Big Bang Nucleosynthesis (BBN), recombination, cosmic microwave background(CMB) and baryon acoustic oscillations should be considered~\cite{Izotov:2014fga,Planck:2018vyg,Riess:2016jrr,Mangano:2011ar}. However, we do not know the temperature of the dark sector after reheating stage. In our study the value of $\Delta N_{\rm eff} <0.5 $ is given and $\frac{T_f}{T_{\rm SM}}=0.47$ is chosen. 
Subsequently, these intersection points in Fig.~\ref{fig2} are used to calculate the
GW signal (with an additional input parameter $h$, as mentioned before). The
signal-to-noise ratios (SNRs) are plotted in Fig.~\ref{fig3}, which shows that in
the future the GW signal produced by the singlet SUSY model may be detected by LISA,
TianQin and Taiji detectors. 

\section{Conclusion}\label{sec5}
We examined  the cosmological phenomena in  
the minimal supersymmetric standard model (MSSM) extended by a singlet superfield, with the coupling between the singlet sector and the MSSM sector being tiny and hence the singlet sector being a quasi dark sector with SUSY. We obtained the following observations: (i) In the parameter space of SIDM which solves the small cosmological scale anomalies,  a FOPT can readily happen but requires rather light dark matter below MeV; (ii) The parameter space required by FOPT and SIDM can be partially covered by detecting the phase-transition GWs at the near-future projects, such as LISA, TianQin and Taiji. Therefore, the recently developed GW astronomy could be a good probe to such a SUSY scenario.

\section*{Acknowledgements}
We thank Bin Zhu and Yang Zhang for useful discussions. This work was supported by the Natural Science Foundation of China under 
grant numbers 11775012, 12075300 and 11821505, by Peng-Huan-Wu Theoretical Physics Innovation Center (12047503), by the CAS Center for Excellence in Particle Physics (CCEPP), by Key R\&D Program of Ministry of Science and Technology of China
under number 2017YFA0402204, and by Key Research Program of the Chinese 
Academy of Sciences under grant No. XDPB15.

\appendix
\section{Feynman rules in the singlet SUSY model}
The interaction vertices are given by 
\be\label{FeynmanRules}\begin{split}
V_{HHH}=&~-2\sqrt{2}\kappa (6\kappa v_s +A_{\kappa}+3\mu_s),\\
V_{HAA}=&~-2\sqrt{2}\kappa (2\kappa v_s -A_{\kappa}+\mu_s),\\
V_{H\chi\chi}=&~-\sqrt{2}\kappa,\\
V_{A\chi\chi}=&~\sqrt{2}i\kappa \gamma^{5}.
\end{split}\ee
The amplitudes for the singlino DM annihilation processes are given by 
\begin{enumerate}
	\item $\chi\chi\to HH$ :
	\bea
	A(^3P_0) &=& 2\sqrt{6}v\kappa^2\left[\frac{R(3m_\chi+A_\kappa)}{4-R(m_H)^2+iG_h}
	-2\frac{1+R(m_\chi)}{P_\chi}+\frac{4}{3}\frac{\bar\beta_f^2}{P_\chi^2}\right] ,\\
	A(^3P_2) &=& -(16/\sqrt{3})v\kappa^2\bar\beta_f^2 /P_\chi^2~.
	\eea
	\item   $\chi\chi\to AA$ :
	\bea
	A(^3P_0) &=& 2\sqrt{6}v\kappa^2\left[\frac{R(m_\chi-A_\kappa)}{4-R(m_H)^2+iG_h}
	-2\frac{1-R(m_\chi)}{P_\chi}+\frac{4}{3}\frac{\bar\beta_f^2}{P_\chi^2}\right], \\
	A(^3P_2) &=& -(16/\sqrt{3}) v\kappa^2\bar\beta_f^2 /P_\chi^2~.
	\eea
	\item   $\chi\chi\to HA$ :
	\bea
	A(^1S_0) &=& -4\sqrt{2}\kappa^2\frac{R(m_\chi-A_\kappa)}{4-R(m_H)^2}\left(1+\frac{v^2}{8}\right)
	\nn\\
	&& +8\sqrt{2}\kappa^2\frac{R(m_\chi)}{P_\chi}\left[1+v^2\left(\frac{1}{8}
	-\frac{1}{2P_\chi}+\frac{\bar\beta_f^2}{3P_\chi^2}\right)\right]\nn\\
	&& +2\sqrt{2}\kappa^2 \left(R(m_A)^2-R(m_H)^2\right)\left[1+v^2\left(-\frac{1}{8}
	-\frac{1}{2P_\chi}+\frac{\bar\beta_f^2}{3P_\chi^2}\right)\right], \\
	A(^3P_1) &=& 8v\kappa^2 \bar\beta_f^2 /P_\chi^2~,
	\eea
\end{enumerate}	

where
\bea
R(m_X)=\frac{m_X}{m_\chi},\quad P_j=1+R(m_j)^2-\frac{1}{2}(R(m_X)^2+R(m_Y)^2),
\quad G_i=\frac{\Gamma_i m_i}{m_\chi^2}.
\eea

\bibliographystyle{apsrev}
\bibliography{note}

\end{document}